\documentclass{amsart}

\usepackage{url}
\usepackage{moreverb}
\usepackage{amsfonts,amssymb,amsbsy,amsmath}
\usepackage{amsaddr}
\usepackage[authoryear]{natbib}
\usepackage[pdftex]{graphicx}

\begin{document}

\title{Probabilistic temperature forecasting based on an ensemble AR modification}

\author{Annette M\"{o}ller}
\address{Department of Animal Sciences, Biometrics \& Bioinformatics Group, University of G\"{o}ttingen, Germany}
\email{annette.moeller@agr.uni-goettingen.de}
\thanks{Corresponding author: Annette M\"{o}ller, Department of Animal Sciences, Biometrics \& Bioinformatics Group, University of G\"{o}ttingen, Carl-Sprengel-Weg 1, D-37075 G\"{o}ttingen.}

\author{J\"{u}rgen Gro{\ss}}
\address{Institute for Mathematical Stochastics, Faculty of Mathematics, Otto von Guericke University Magdeburg, Germany}
\email{juergen.gross@ovgu.de}


\keywords{ensemble postprocessing; predictive probability distribution; autoregressive process; spread-adjusted linear pool}

\date{}

\begin{abstract}
To address the uncertainty in outputs of numerical weather prediction (NWP) models, ensembles of forecasts are used. To obtain such an ensemble of forecasts the NWP model is run multiple times, each time with different formulations and/or initial or boundary conditions. To correct for possible biases and dispersion errors in the ensemble, statistical postprocessing models are frequently employed. These statistical models yield full predictive probability distributions for a weather quantity of interest and thus allow for a more accurate assessment of forecast uncertainty.
This paper proposes to combine the state of the art Ensemble Model Output Statistics (EMOS) with an ensemble that is adjusted by an AR process fitted to the respective error series by a spread-adjusted linear pool (SLP) in case of temperature forecasts.
The basic ensemble modification technique we introduce may be used to simply adjust the ensemble itself as well as to obtain a full predictive distribution
for the weather quantity. As demonstrated for temperature forecasts of the
European Centre for Medium-Range Weather Forecasts
(ECMWF) ensemble, the proposed procedure gives rise to improved results upon the basic
(local) EMOS method.
\end{abstract}

\maketitle

\markboth{A.~M\"{o}ller, J.~Gro{\ss}}{Ensemble AR modification}

\section{Introduction}\label{sec:intro}

Weather forecasting is usually based on outputs of numerical weather prediction (NWP) models that represent the dynamical and physical behaviour of the atmosphere. Based on actual  weather conditions, the equations are employed to extrapolate the state of the atmosphere, but may strongly depend on initial conditions and other uncertainties of the NWP model.
A methodology that accounts for such shortcomings is the use of ensemble forecasts by running the NWP model several times with different initial conditions and/or model formulations \citep{gneiting2005calibrated, LeutbecherPalmer2008}.

Forecast ensembles play an important role when it desired to develop methods that transfer from deterministic to probabilistic forecasting, since information about forecast mean and variance may be extracted from several individual forecasts \citep{Palmer2002}. In practise, however, ensemble prediction systems are not able to capture all sources of uncertainty, thus they often exhibit dispersion errors (underdispersion) and biases. Statistical postprocessing models correct the forecasts in accordance with recent forecast errors and observations and yield full predictive probability distributions \citep[see e.g.][]{GneitingRaftery2005, WilksHamill2007, gneiting2014probabilistic}.

There are two widely used approaches in statistical post\-processing that yield full predictive distributions based on a forecast ensemble and verifying observations. In the Bayesian model averaging (BMA, \citealp{Raftery&2005}) approach each ensemble member is associated with a kernel density (after suitable bias correction) and the individual densities are combined in a mixture distribution with weights that express the skill of the individual ensemble members. Ensemble model output statistics (EMOS, \citealp{gneiting2005calibrated}) combines the ensemble members in a multiple linear regression approach with a single predictive distribution.
\citet{gneiting2005calibrated} and \citet{Raftery&2005} apply the postprocessing methods to weather quantities, where a normal distribution can be assumed as underlying model, such as temperature and pressure. For the application to other weather quantities alternative distributions are required. An overview on existing variants of EMOS and BMA can for example be found in \citet{Schefzik&2013} and \citet{gneiting2014probabilistic}.

In cases where various probabilistic forecasts from different sources are available, combining these forecasts can improve the predictive ability further. The individual forecasts might for example (as in our application) come from different competing statistical postprocessing models. The most widely used method to combine the individual predictive distributions is the linear pool (LP), see for example \citet{gneiting2013combining}, \citet{RanjanGneiting2010} and references therein for reviews on the topic.
\citet{gneiting2013combining} show that the linear pool results in an overdispersed predictive distribution, regardless whether the individual components are calibrated or not.
A more flexible and non-linear approach is the spread-adjusted linear pool (SLP).
For example, \citet{Berrocal&2007} and \citet{Kleiber&a2011} empirically observed overdispersion of the linear combined forecasts in case of approximately neutrally dispersed Gaussian components and they introduced a non-linear spread-adjusted combination approach, that was generalized and discussed by \citet{gneiting2013combining}.
Further, \citet{gneiting2013combining} and \citet{RanjanGneiting2010} propose another flexible non-linear aggregation method, the Beta-transformed linear pool (BLP), resulting in highly improved dispersion properties.

Although weather prediction implies temporal structures and dependencies, we were unable to find explicit applications of time series methods in the context of ensemble postprocessing methods. On the contrary, there are several approaches to apply time series models directly on observations of a weather quantity or other environmental and climatological quantities.
To name only a few, \citet{brown1984time} apply an $\text{AR}(p)$ model to transformed wind speed data in order to simulate and predict wind speed and wind power observations. \citet{KatzSkaggs1981} investigate the problems when fitting ARMA models to meteorological time series and, as an example, present an application to time series of the Palmer Drought index. \citet{AlAwadhiJolliffe1998} fit specific ARMA models to surface pressure data. \citet{MilionisDavies1994} apply the Box-Jenkins modelling technique to monthly activity of temperature inversions.

We propose to apply time series models in the context of ensemble postprocessing.
To account for an autoregressive (AR) structure in time we construct a predictive distribution based on an AR-adjusted forecast ensemble (local AR-EMOS). As the standard local EMOS predictive distribution shows signs of underdispersion and our local AR-EMOS distribution on the contrary clearly exhibits overdispersion we propose to combine AR-EMOS and EMOS
with a spread-adjusted linear pool.

The structure of the paper is the following: Section \ref{sec:methods} presents our proposed AR modification of the forecast ensemble and briefly reviews the EMOS model and the spread-adjusted linear combination of probabilistic forecasts. Sections \ref{sec:applicationtemp} and \ref{sec:applicationsingle} illustrate some possible applications of our basic AR-modification method in a case study with temperature forecasts of the European Center for Medium Range Weather Forecasts (ECMWF).
We end the paper with some concluding remarks and a discussion of further extensions of our method.

\section{Methods} \label{sec:methods}

\subsection{Ensemble model output statistics} \label{sec:emos}

\citet{gneiting2005calibrated} introduced the ensemble model output statistics (EMOS) model for the case, that a normal distribution can be assumed as model for the weather quantity.
In all following sections we use a notation depending on the time index $t$, to stress the fact that we explicitly model temporal dependencies through our autoregressive adjustment approach.

The predictive distribution is obtained by fitting the following multiple linear regression model to the observation $y(t)$ of the weather quantity $Y(t)$ and the forecast ensemble $\{X_{1}(t),\ldots, X_{m}(t)\}$:
\begin{equation}\label{emos1}
Y(t) = a + b_1 X_1(t) + \ldots + b_m X_m(t) + \varepsilon(t),
\end{equation}
where $a, b_1, \ldots, b_m \in \mathbb{R}$ are real valued regression coefficients that can be interpreted as bias-correction coefficients, and $\varepsilon(t)$ is a normally distributed error term with $\varepsilon(t) \sim \mathcal{N}(0, \sigma^2(t))$. For convenience we define $\xi(t)=a + b_1 X_1(t) + \ldots + b_m X_m(t)$, yielding the representation
\begin{equation}
Y(t) = \xi(t) + \varepsilon(t)
\end{equation}
of the EMOS model.

In case of an exchangeable ensemble, the multiplicative bias correction parameters are chosen to be equal, that is $b_i=b$, for $i=1,\ldots,m$. In this case, the EMOS model is given as
\begin{equation}\label{emos11}
Y(t) = a + b \overline{X}(t) + \varepsilon(t),
\end{equation}
where $\overline{X}(t) = \sum_{i=1}^m X_i(t)/m$ and we can define $\xi(t)=a + b \overline{X}(t)$.

The variance of $\varepsilon(t)$ is parameterized as linear function of the ensemble variance to account for dispersion errors in the raw ensemble:
\begin{equation} \label{varemos}
\textup{Var}({\varepsilon(t)}) = \sigma^2(t) = c + d S^2(t),
\end{equation}
where $S^2(t) = \sum_{i=1}^m (X_i(t) - \overline{X}(t))^2/(m-1)$ and $c, d \in \mathbb{R}_+$ are nonnegative coefficients.

Models \eqref{emos1} and \eqref{emos11} both yield the following predictive EMOS distribution:
\begin{equation} \label{emos2}
Y(t)|X_1(t),\ldots,X_m(t) \sim \mathcal{N}\Big(\xi(t), \sigma^2(t) \Big).
\end{equation}
The parameters of the distribution are estimated from a training period preceding the forecast time and re-estimated for each forecast time using a rolling training period of fixed length. \cite{gneiting2005calibrated} propose to estimate the parameters by optimizing the continuous ranked probability score (CRPS, \citealp{wilks2011statistical}) over the training data. This estimation procedure is implemented in the {\sf R} package {\tt ensembleMOS} \citep{R2015,ensembleMOS}.

Two variants of estimating the EMOS parameters exist, the global and the local EMOS approach. For global EMOS only one set of parameters is estimated, while for global EMOS a separate set is estimated at each station.

For our case study we consider a station-wise approach and therefore employ the local EMOS variant.

\begin{figure}
\centering
\includegraphics[width=20pc,angle=0]{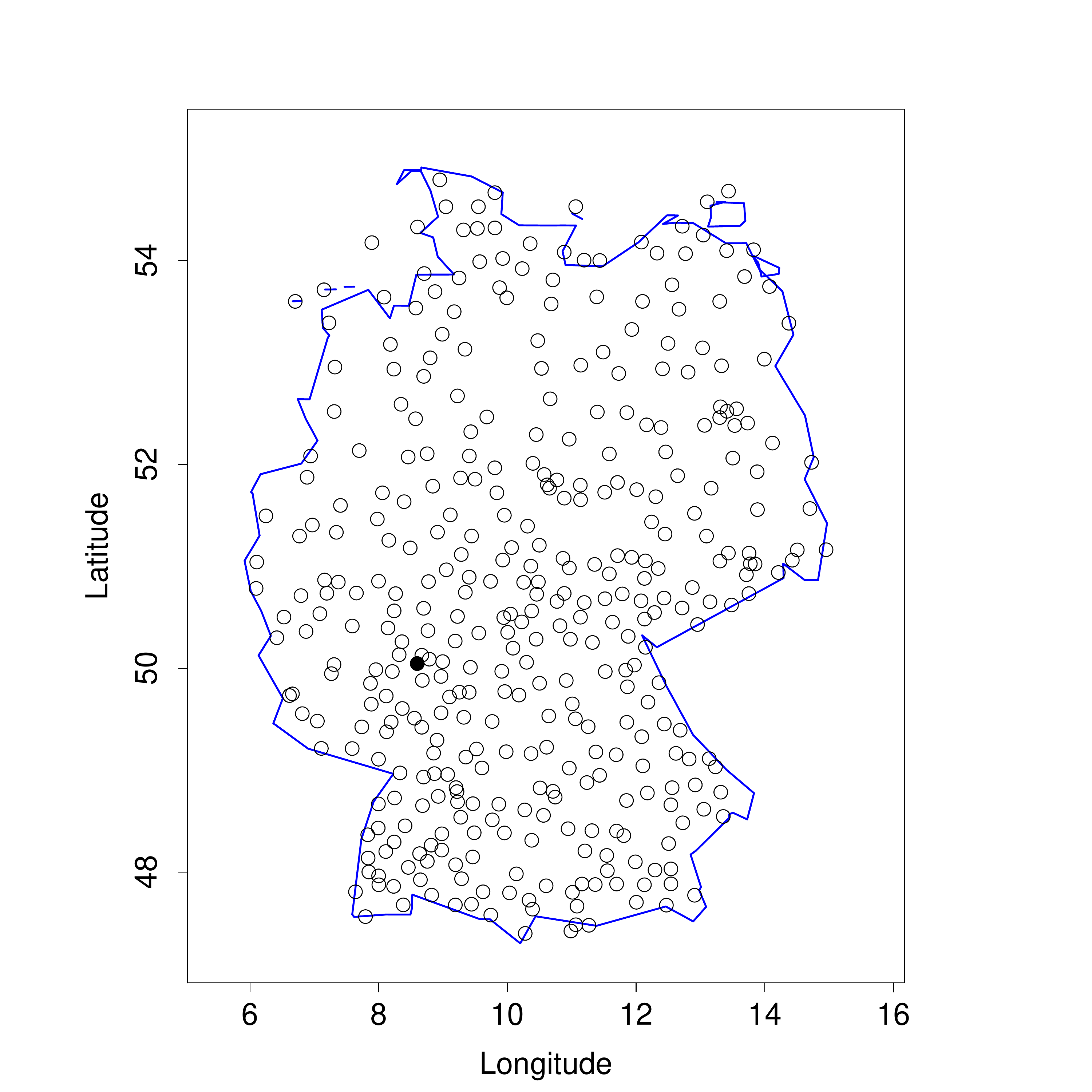}\\
\caption{The 383 observation stations in Germany used for the analysis, where the filled circle indicates station Frankfurt a.M.}\label{fstations}
\end{figure}

\subsection{The AR modification technique}\label{sec:ar}

Let again $\{X_{1}(t),\ldots, X_{m}(t)\}$ denote an ensemble of
forecasts for a univariate (normally distributed) weather quantity $Y(t)$ at a fixed location. Let $\eta(t)$ denote
a deterministic-style forecast of $Y(t)$ with corresponding forecast error
\begin{equation}\label{E1}
Z(t) := Y(t) - \eta(t)\; .
\end{equation}
Simple examples for $\eta(t)$ are
provided by any ensemble member $X_{i}(t)$ itself, the raw ensemble mean $\overline{X}(t) = \sum_{i=1}^{m} X_{i}(t)/m$, or the raw ensemble median denoted by $\dot{X}$.

If $\eta(t)$ is a one-step-ahead forecast made at origin $t-1$, then $\eta(t)$
may be seen as appropriate, if the time series $\{Z(t)\}$ can be viewed as a white noise process.
If there is some indication that this property is violated, one may readily
assume that the series $\{Z(t)\}$ follows a weakly stationary
AR$(p)$ process, i.e.
\begin{equation}\label{E2}
Z(t) - \mu = \sum_{j=1}^{p} \alpha_{j}[Z(t-j)- \mu] + \varepsilon(t)\; ,
\end{equation}
where $\{\varepsilon_{t}\}$ is white noise. Combing (\ref{E1}) and (\ref{E2})
gives $Y(t) = \widetilde{\eta}(t)  + \varepsilon(t)$, where
\begin{equation}\label{E3}
\widetilde{\eta}(t) = \eta(t) + \mu + \sum_{j=1}^{p} \alpha_{j}[Y(t-j) - \eta(t-j) - \mu]
\end{equation}
can be seen as an AR modified forecast based on the actual forecast $\eta(t)$ and past values $Y(t-j)$ and $ \eta(t-j)$, $j=1,\ldots, p$. The coefficients $\mu$, $\alpha_{1}$, \ldots , $\alpha_{p}$ may be obtained by
fitting an $\text{AR}(p)$ process to the observed error series $\{Z(t)\}$ from a
training period, where the order $p$ of the process can automatically be chosen by applying an appropriate criterion. This includes the incidence $p=0$, in which case
$\widetilde{\eta}(t)$ is a simple bias correction of $\eta(t)$.
For the actual fitting we employ Yule-Walker estimation as carried out by
the {\sf R} function {\tt ar}, see also \citet[Section 3.6]{shumway2006time}.
Order selection is done by the AIC criterion, despite the circumstance that the estimated coefficients are not the maximum likelihood estimates.

The described approach differs from the times series methods
introduced e.g. by \citet{brown1984time} in the sense that it does {\em not} aim at directly modelling
a weather quantity itself.

The basic AR modification method can be employed for different purposes in ensemble postprocessing, depending on the need of the user. For example it can simply be used to obtain an AR-modified raw ensemble, or to build a postprocessed predictive distribution based on the modified ensemble.
The following sections illustrate some possible applications of AR modified ensemble
forecasts by means of a given data set, where we analyze the aggregated predictive performance and the performance at a single station.
An AR-EMOS predictive distribution for temperature is introduced and shown to
improve upon the well-established local EMOS method with respect to certain verification scores.

See e.g. \citet{wilks2011statistical} as a reference for
applied statistical methods for atmospheric sciences
and \citet{gneiting2014probabilistic}
for a comprehensive review of probabilistic forecasting.

\subsection{Spread-adjusted linear combination of predictive distributions} \label{sec:linpool}

Let $\mathcal{D}_{\mathbb{R}}^+$ denote the class of non-random cumulative distribution functions (CDFs) that admit a Lebesque density, have support on $\mathbb{R}$ and are strictly increasing on $\mathbb{R}$. Further, $F_1,\ldots,F_k \in \mathcal{D}_{\mathbb{R}}^+$ denote the considered CDFs with Lebesgue densities $f_1,\ldots,f_k$, where $k$ is an arbitrary but finite integer value.

Then the spread-adjusted linear pool (SLP) combined predictive distribution with spread-parameter $c$ has CDF (see \citealp{gneiting2013combining})
\begin{equation} \label{slpcdf}
G_c(y) = \sum_{l=1}^k \; w_l \, F_l^0 \Big( \frac{y - \mu_l}{c} \Big)
\end{equation}
and density
\begin{equation} \label{slppdf}
g_c(y)= \frac{1}{c} \, \sum_{l=1}^k \; w_l \, f_l^0 \Big( \frac{y - \mu_l}{c} \Big),
\end{equation}
where $w_1,\ldots,w_k$ are non-negative weights with $\sum_{l=1}^k \, w_l = 1$, $c>0$ is the spread-adjustment parameter and $\mu_l$ is the unique median of $F_l$. Further $F_l^0$ and $f_l^0$ are defined via the relationship $F_l(y)=F_l^0(y - \mu_l)$ and $f_l(y)=f_l^0(y - \mu_l)$, respectively.

\citet{gneiting2013combining} note that for neutrally dispersed or overdispersed components, a value of $c<1$ may be appropriate, while for underdispersed components, a value $c\ge1$ is suggested. A value of $c=1$ corresponds to the standard linear pool.

Typically a common spread parameter is used for all components, although the method can be generalized to have spread parameters varying with the components. However, this may be appropriate only in case the dispersion properties of the components differ to a high extent.

\begin{figure}
\centering
\includegraphics[width=20pc,angle=0]{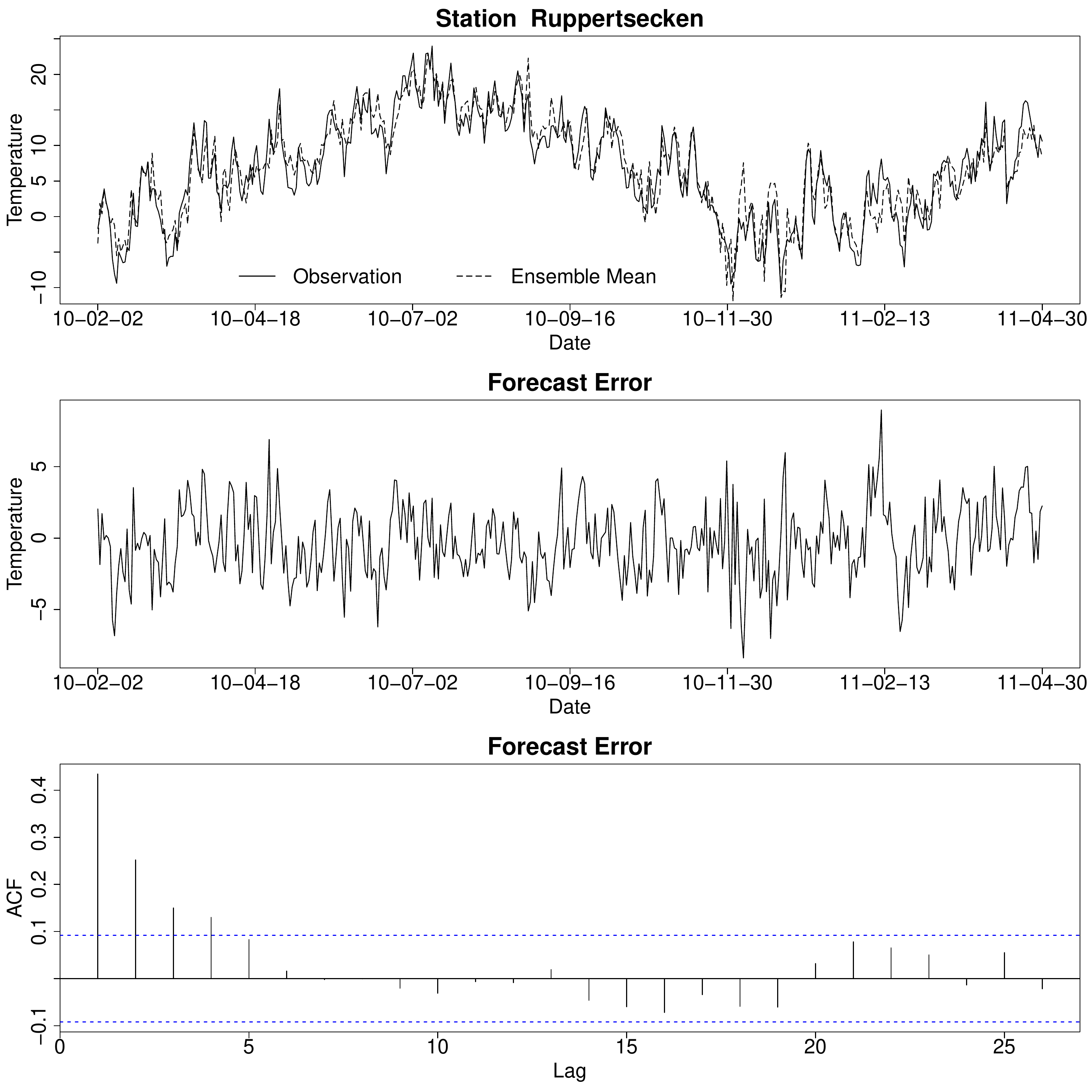}\\
  \caption{Series of temperature and ensemble mean (upper panel), series of forecast errors (middle panel), and ACF of series of forecast errors (lower panel) for station Ruppertsecken}\label{f3}
\end{figure}

\section{Application to European Center for Medium Range Weather Forecasts of temperature} \label{sec:applicationtemp}

The data set for our case study comprises an ensemble with $m=50$ members (and one control forecast not used here) of the European Centre for Medium-Range Weather Forecasts (ECMWF, see e.g. \citealp{molteni1996ecmwf}). Initialized at 00 UTC, the forecasts are issued on a grid with 31 km resolution, and they are available for forecast horizons in 3 hour steps up to 144 hours. In Germany 00 UTC corresponds to 1am local time, and to 2am local time during the daylight saving period.
For our analysis we consider $24$-h
ahead forecasts for 2-m surface temperature in Germany along with the verifying observations at
different stations in the time period ranging from 2010-02-02 to 2011-04-30.
Although there is a total of 518 stations in the full data set, only $383$ stations with complete $T=453$
observations for the variables were retained.
To use the forecasts in combination with globally distributed surface synoptic observations (SYNOP) data, they are bilinearly interpolated from the four surrounding grid points to the locations that correspond to actual observation stations. The observation data from stations in Germany were provided by the German Weather Service (DWD).
Figure \ref{fstations} shows the locations of the 383 stations within Germany, where the station Frankfurt a.M. is marked as a filled circle.

\subsection{Forecast error of ensemble mean}

For each of the 383 stations we compute the series $Z(t) = Y(t) - \overline{X}(t)$
of forecast errors of the ensemble mean, where $t$ ranges over the whole time period.
To check for independence, we apply the Ljung-Box test \citep{ljung1978measure}
based on lag 1, which is available in {\sf R} as function {\tt Box.test}.
All 383 computed p-values are not greater than 0.046 (the largest occurring value), indicating
substantial autocorrelation in the forecast error series for each station. Figure \ref{f3} further illustrates this point by showing the series of temperature observations together with the ensemble mean, the corresponding forecast errors, and the autocorrelation function (ACF) of the series of forecast errors for the randomly chosen station Ruppertsecken in Rhineland-Palatinate.

To account for autocorrelation in a series of forecast errors, the following subsections present an approach to obtain a predictive distribution based on the AR modification method (AR-EMOS), while Section \ref{sec:applicationsingle} discusses AR-EMOS for the station Frankfurt a.M. with regard to a much longer than before verification period of 3650 days.

\subsection{Length of training period for fitting the AR model}

When the mean $\overline{X}(t)$ of the raw ensemble
is considered as a deterministic-style forecast $\eta(t)$, a corresponding AR modified
forecast $\widetilde{\eta}(t)=\widetilde{\overline{X}}(t)$ can be generated according to the procedure
described in Section \ref{sec:ar}.
For computing the order and the coefficients of the AR$(p)$ process,
a training period of $T_{1}$ days previous to the forecast day is considered.
For each day out of the $T_{2} = T - T_{1}$ remaining days from the forecast (verification)
period, the AR coefficients are newly estimated, where the training period is shifted
accordingly (rolling training period).

When choosing an appropriate length of training period, there is usually a trade-off. Longer training periods yield more stable estimates, but may fail to account for temporal changes, while shorter periods can adapt more successfully to changes over time.
To learn about an appropriate training length, a variety of possible values, namely $T_{1}=30, 60, 90, 120, 150, 180, 210$,
were considered in a preliminary analysis, and for each station the mean absolute error
\begin{equation}
\text{MAE} = \frac{1}{T_{2}} \sum_{t=1}^{T_{2}} | Y(t) - \widetilde{\eta}(t)|\; ,
\end{equation}
of the AR modified forecast $\widetilde{\eta}(t) = \widetilde{\overline{X}}(t)$ has been computed.
As it turned out, a training length of $T_{1}= 90$ days provided a favorable
performance with respect to the MAE averaged over the considered stations.
This number of days will be used as training length for fitting the AR process in all further analyses.

\begin{table}
\caption{MAEs for $T_{2}=363$ days averaged over 383 stations.}\label{t1}
\centering
\begin{tabular}{lrrrrrr}
\hline
$\eta(t)$ & $\overline{X}(t)$ & $\widetilde{\overline{X}}(t)$ & $\overline{\widetilde{X}}(t)$ & $\dot{X}(t)$ & $\widetilde{\dot{X}}(t)$ & $\dot{\widetilde{X}}(t)$\\
\hline
MAE & 2.911  & 2.024  & 2.007 & 2.918 & 2.037 & 2.018\\
\hline
\end{tabular}
\end{table}

\subsection{AR modified ensemble} \label{sec:arens}

In a second step the procedure from the previous subsection is applied directly
to each ensemble member.
For each station, this procedure generates an AR modified ensemble for the forecast period of
length $T_{2}$. Table \ref{t1} shows a comparison of MAEs corresponding to certain
deterministic-style forecasts.
Here, $\widetilde{\overline{X}}(t)$ is the AR modification of the raw ensemble mean, while $\overline{\widetilde{X}}(t)$ is the mean of the AR modified ensemble members. For the dot superscript denoting the median, the two definitions are analog.

For the mean as well as the median, the approach of first applying the AR modification to each member of the raw ensemble and then computing the deterministic style forecast from the modified ensemble yields smaller MAE values than the reverse procedure of applying the AR modification directly to the deterministic style forecast computed from the original ensemble.
As it turns out, the mean of the AR modified ensemble $\overline{\widetilde{X}}(t)$
performs best among the considered deterministic-style forecasts with respect to MAE and will therefore be the basis for constructing a predictive probability distribution in the following.

\begin{figure}
\centering
\includegraphics[width=20pc,angle=0]{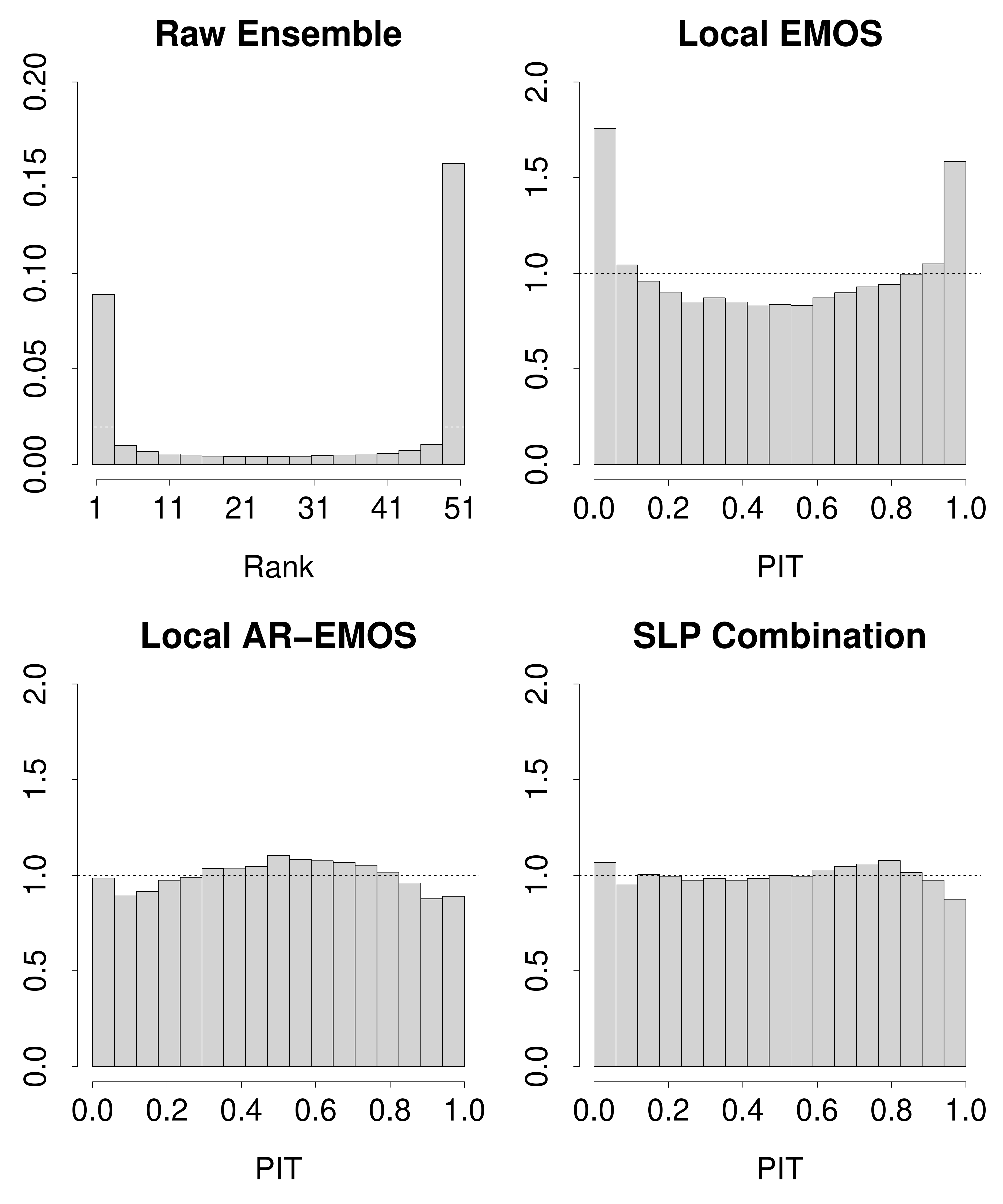}\\
  \caption{Univariate verification rank histogram and PIT histograms over
383 stations and all dates in the verification period.}\label{f1}
\end{figure}

\subsection{Predictive distribution} \label{sec:preddistoverall}

The above may be seen as a preliminary to the development of a statistical
postprocessing method yielding a full predictive distribution, based on the AR modification. The EMOS procedure described by \citet{gneiting2005calibrated} and shortly reviewed in Section \ref{sec:emos} assumes a
Gaussian predictive distribution
\begin{equation}
\mathcal{N}(\xi(t), \sigma^2(t))\; ,
\end{equation}
for a weather quantity $Y(t)$, given the ensemble forecasts $(X_1(t), \ldots, X_m(t))$.
Here, $\xi(t)$ is a linear combination
of the ensemble members and $\sigma^2(t)$ is a linear function of the ensemble variance, as described in Equations \eqref{emos1}, \eqref{emos11} (exchangeable ensemble version) and \eqref{varemos}.
The coefficients are obtained station-wise (local EMOS) by a minimum-CRPS estimation procedure.

To obtain a predictive distribution based on our AR modification (AR-EMOS) we employ a plug-in strategy replacing $\xi(t)$ by the mean of the
AR modified ensemble $\overline{\widetilde{X}}(t)$ from step 2 outlined in Section \ref{sec:arens}.
When the deterministic-style forecast $\eta(t)$ in \eqref{E1} is considered as non-random,
\begin{equation}
\textup{Var}(Y(t)) = \textup{Var}(Z(t)) = \sigma_{\varepsilon}^2 \left(1+ \sum_{j=1}^{\infty} \psi_{j}^2\right)\; ,
\end{equation}
where $\sigma_{\varepsilon}^2$ is the variance of $\varepsilon(t)$ and the $\psi_{j}$ are the coefficients of the one-sided linear process (moving average) representation of $Z(t)$, cp. \citet[Section 3.2]{shumway2006time}.

We obtain the $\psi_{j}$ coefficients by replacing the unknown AR coefficients by their estimates and solve a homogenous difference equation, see. e.g.
\citet[Example 3.10]{shumway2006time}. In {\sf R} this can be done by the function {\tt ARMAtoMA}. By replacing $\sigma_{\varepsilon}^2$ by its estimate and using
only the first ten $\psi_{j}$-weights, we compute an estimate of $\textup{Var}(Y(t))$ for every
ensemble member $\eta(t) = X_{i}(t)$. The Gaussian parameter $\sigma^2(t)$ is then
estimated as the average variance estimate.

The local EMOS procedure itself requires an additional training period. Preliminary investigations suggested that a length of 25 days is optimal for the data set of our case study.
The training period for fitting the AR process together with the training period for fitting the EMOS model yields a forecast period of $T_{2} = 338$ days (ranging from 2010-05-28 to 2011-04-30) for
verification purposes.

The MAE of $\overline{\widetilde{X}}(t)$ averaged over the stations
in this final verification period is $2.036$, and thus slightly smaller than the corresponding
averaged MAE $2.042$ of the local EMOS predictive mean, see Table \ref{t2}.
By considering the station averaged CRPS, the local EMOS predictive distribution admits a
value of $1.471$ while the CRPS of the
local AR-EMOS distribution shows a slightly better value of $1.460$.
The DSS score (see Section \ref{sec:slpoverall}) of local AR-EMOS is smaller than the value of local EMOS as well.
A comparison of the rank histogram
for the raw ensemble, and probability integral transform (PIT) histograms for the local EMOS
method and the local AR-EMOS displayed in Figure \ref{f1}
indicates a good calibration of the AR modification.
However, while local EMOS still shows signs of underdispersion, the small hump shape of
local AR-EMOS reveals a slight tendency to the reverse effect (overdispersion). These contradictory dispersion properties suggest
a combination of both predictive distributions. The combination with a spread-adjusted linear pool is discussed in the next section and reveals an
improvement in calibration compared
the individual predictive distributions.

\begin{table}
\caption{Verification statistics averaged over $T_{2}=338$ days
and 383 stations.}\label{t2}
\centering
\begin{tabular}{lrrrrrr}
\hline
~ & EMOS & AR-EMOS & SLP \\
\hline
MAE &  2.042 & 2.036 & 1.969\\
CRPS & 1.471 & 1.460 & 1.407\\
DSS & 3.135 &  2.908 & 2.821\\
\hline
\end{tabular}
\end{table}

\subsection{Combination of predictive distributions} \label{sec:slpoverall}

In this subsection we introduce a spread-adjusted combination of the local EMOS and the local AR-EMOS predictive distributions obtained in the previous section. We perform the aggregation of the distributions in line with the spread-adjusted linear pool (SLP) formula of \citet{gneiting2013combining}, that we shortly reviewed in Section \ref{sec:linpool}.

We define the SLP combined cumulative distribution function according to Equation \eqref{slpcdf} with $k=2$ in our case.
Now, let $\phi$ and $\Phi$ denote the probability density function (PDF) and the
cumulative distribution function (CDF)
of the standard normal distribution, respectively.
If the two component CDFs we wish to combine are normal, then
$F_{l}^{0}(y) = \Phi(y/\sigma_{l})$, $l=1,2$, in Eq. (\ref{slpcdf}),
and the SLP combined predictive CDF is
\begin{equation}\label{E4}
F(y) = w_{1} G_{1}(y) + w_{2} G_{2}(y), \quad G_{l}(y) =
\Phi\left(\frac{y- \mu_{l}}{\sigma_{l} c}\right)\; ,
\end{equation}
$l=1,2$, where $w_{1}$ is nonnegative, $w_{2}= 1- w_{1}$, and $c$ is a strictly positive
spread adjustment parameter \citep{gneiting2013combining}. The  choice $c=1$
corresponds to the traditional linear pool (TLP). The expectation $\mu_{F}$ and the
variance $\sigma_{F}^2$ of the CDF $F$ are
\begin{equation}
\mu_{F} = \sum_{l=1}^{2} w_{l} \mu_{l}
\end{equation}
and
\begin{equation}
\sigma_{F}^2 = \sum_{l=1}^{2} w_{l} (\mu_{l}^2 + c^2 \sigma_{l}^2) - \mu_{F}^2\; ,
\end{equation}
allowing for straightforward computation of the \citet{dawid1999coherent} score
\begin{equation}
\text{DSS}(F, y_{\text{obs}}) = \frac{(y_{\text{obs}} - \mu_{F})^2}{\sigma_{F}^2} + 2\log \sigma_{F}\; ,
\end{equation}
see also \citet{gneiting2014probabilistic}.

From Equation (5) in
\citet{grimit2006continuous}, the CRPS
\begin{equation}\label{E10}
\text{CRPS}(F, y_{\text{obs}}) = \int_{-\infty}^{\infty}
\left\{F(y) - \boldsymbol{1}(y\geq y_{\text{obs}})\right\}^{2}
\, \text{d} y\; ,
\end{equation}
where $F$ is of the form \eqref{E4}, can also be written as
\begin{equation}\label{E9}
\begin{split}
\text{CRPS}(F, & y_{\text{obs}}) =
\sum_{l=1}^{2} w_{l}  A(y_{\text{obs}} - \mu_{l}, c^2 \sigma_{l}^2)\\
& - \frac{1}{2} \sum_{l=1}^{2}\sum_{k=1}^{2} w_{l} w_{k} A(\mu_{l} -\mu_{k}, c^2(
\sigma_{l}^2 + \sigma_{k}^2))
\\
\end{split}
\end{equation}
where
\begin{equation}
A(\mu,\sigma^2) = 2\sigma\phi\left(\frac{\mu}{\sigma}\right) + \mu\left(2
\Phi\left(\frac{\mu}{\sigma}\right) - 1\right)
\end{equation}
is the expectation of the absolute value of a $\mathcal{N}(\mu, \sigma^2)$ distributed
random variable. Since there exist approximations of $\Phi$ up to an arbitrary precision,
see e.g. \citet{abramowitz1972handbook}, Formula \eqref{E9} is easy to evaluate.

\subsubsection{Choice of appropriate weights} \label{sec:slpweights}

To gain some insight about an appropriate choice of the SLP parameters when combining local EMOS and local AR-EMOS,
we investigate a grid of combinations of values for $w_1$ ($w_2$ is fully determined by $w_1$) and $c$.
More precisely, we obtain the SLP combined predictive distribution for all 99 combinations
$w_{1} =0.0$, $0.1$, \ldots, $0.9$, $1.0$ ($w_{2} = 1 - w_{1}$) and
$c = 0.6, 0.7, \ldots 1.3, 1.4$.

For each of the above mentioned combinations, the
average DSS and CRPS over $T_{2}=338$ days and 383 stations is computed. As it turns out,
the minimal average DSS and CRPS values both occur for the simple unfocused combination
$w_{1} = w_{2} =0.5$ and $c=1$. As it seems, within the unfocused SLP combination the contradictory dispersion properties of local EMOS and local AR-EMOS mutually compensate, yielding a predictive distribution with improved calibration.
To this effect, the accordingly combined predictive CDF is
favourable upon both local EMOS and local AR-EMOS with respect
to the DSS and the CRPS, and
performs better with respect to the MAE as well, see Table \ref{t2}.
Figure \ref{f1} shows the verification rank histogram of the raw ensemble and the PIT histogram of the three considered postprocessing methods. The rank of the raw ensemble (top right panel) exhibits a very pronounced U-shape indicating a strong underdispersion and a need for postprocessing.
The PIT histogram of the SLP combination, displayed in the bottom right panel of Figure \ref{f1}, shows highly improved dispersion properties. While local EMOS (top right panel) clearly exhibits a U-shape and local AR-EMOS (bottom left panel) a slight hump-shape, the SLP combination is corrected for both types of dispersion errors, resulting in PIT histogram that is close to uniformity.

\begin{table}
\caption{PIT variance (dispersion) and density forecast root mean variance (sharpness)
for $383$ stations and $338$ verfications days.}\label{t4}
\centering
\begin{tabular}{lll}
\hline
& $\text{Var(PIT)}$ & $\text{RMV}$ \\
\hline
Local EMOS & 0.101& 2.30\\
Local AR-EMOS & 0.079 & 2.73\\
SLP Combination ($w_{1} =0.5, c=1.0$) &0.083& 2.62\\
\hline
\end{tabular}
\end{table}

The improved dispersion properties visible in the PIT histogram are further supported by Table \ref{t4} showing the variance of the PIT values and the root of the mean variance of the respective postprocessing methods. A similar table was considered in \citet{gneiting2013combining} to investigate the proposed types of combinations of predictive distributions. The variance of the PIT values provides further information on the dispersion properties of the distribution, a variance equal to $\frac{1}{12}=0.0833$ corresponds to the variance of the uniform distribution on $[0,1]$, thus indicating neutral dispersion \citep{gneiting2013combining}. The root mean variance is used as a sharpness measure of the predictive distributions. A main principle of probabilistic forecasting is ``maximizing the sharpness of the predictive distribution subject to calibration'' \citep[see e.g.][]{gneiting2014probabilistic}, therefore the sharpness should be investigated in conjunction with the calibration.
When looking at  Table \ref{t4} we can see that the PIT values of local EMOS have a variance $> \frac{1}{12}$, indicating underdispersion, while for the PIT values of local AR-EMOS we have a variance slightly smaller than $\frac{1}{12}$, indicating overdispersion \citep{gneiting2013combining}. The variance of the PIT values of our derived SLP combination is virtually identical to $\frac{1}{12}$. These results are in line with the PIT histograms in Figure \ref{f1}. When investigating the sharpness in terms of the root mean variance, we see that the local EMOS predictive distribution is the sharpest, while local AR-EMOS the least sharpest one. The sharpness of our SLP combination lies between the other two. Although local EMOS exhibits the highest sharpness, it strongly lacks calibration, leading to a deterioration in predictive performance. On the contrary, local AR-EMOS shows a decreased sharpness, but is uncalibrated as well. In view of the sharpness principle mentioned above, our SLP combination is able to combine improvement in sharpness and calibration in comparison to both, local EMOS and local AR-EMOS.

\subsubsection{Testing for forecast accuracy} \label{sec:dmtest}

The improvement in the verification scores of our derived SLP combination
over the local EMOS method may also be investigated for significance by testing for equal predictive
performance of the two methods with the Diebold-Mariano test for time series, see \citet{gneiting2014probabilistic}.

Let $g_{t}^{(j)}$, $j=1,2$, denote the series of mean CRPS values (averaged over all 383 stations) of local EMOS and
our unfocused SLP combination derived in Section \ref{sec:slpweights} for the verification period of length $T_{2} = 338$,
respectively. Then the large-sample standard normal test statistic adapted from
\citet{diebold1995comparing} is
\begin{equation}
S = \sqrt{T_{2}} \frac{\overline{d}}{\sqrt{\sum_{\tau = - (h-1)}^{h-1}
\widehat{\gamma}_{d}(\tau)}}\; ,
\end{equation}
where
\begin{equation}
\overline{d} = \frac{1}{T_{2}} \sum_{t=1}^{T_{2}} d_{t}, \quad
d_{t} = g_{t}^{(1)} - g_{t}^{(2)} \; ,
\end{equation}
is the average CRPS differential and
\begin{equation}
\widehat{\gamma}_{d}(\tau) = \frac{1}{T_{2}}
\sum_{t=|\tau|+1}^{T_{2}} \, (d_{t} - \overline{d})(d_{t-|\tau|} - \overline{d})
\end{equation}
are the empirical autocovariances.

Usually the truncation lag $h-1$ refers to $h$-step
ahead forecast errors, suggesting the choice $h=1$ here.

When comparing the differential series of local EMOS and the combined distribution for the case $h=1$, yields
$S=7.0621$, which, compared to the standard normal distribution,
gives a clear indication of nonequal predictive performance of the two methods in question
and therefore of improvement of our above derived SLP combination upon local EMOS.
Higher choices of $h$ yielded different values of the test statistic $S$, all of which were
found as highly significant.

\begin{figure}
\centering
\includegraphics[width=20pc,angle=0]{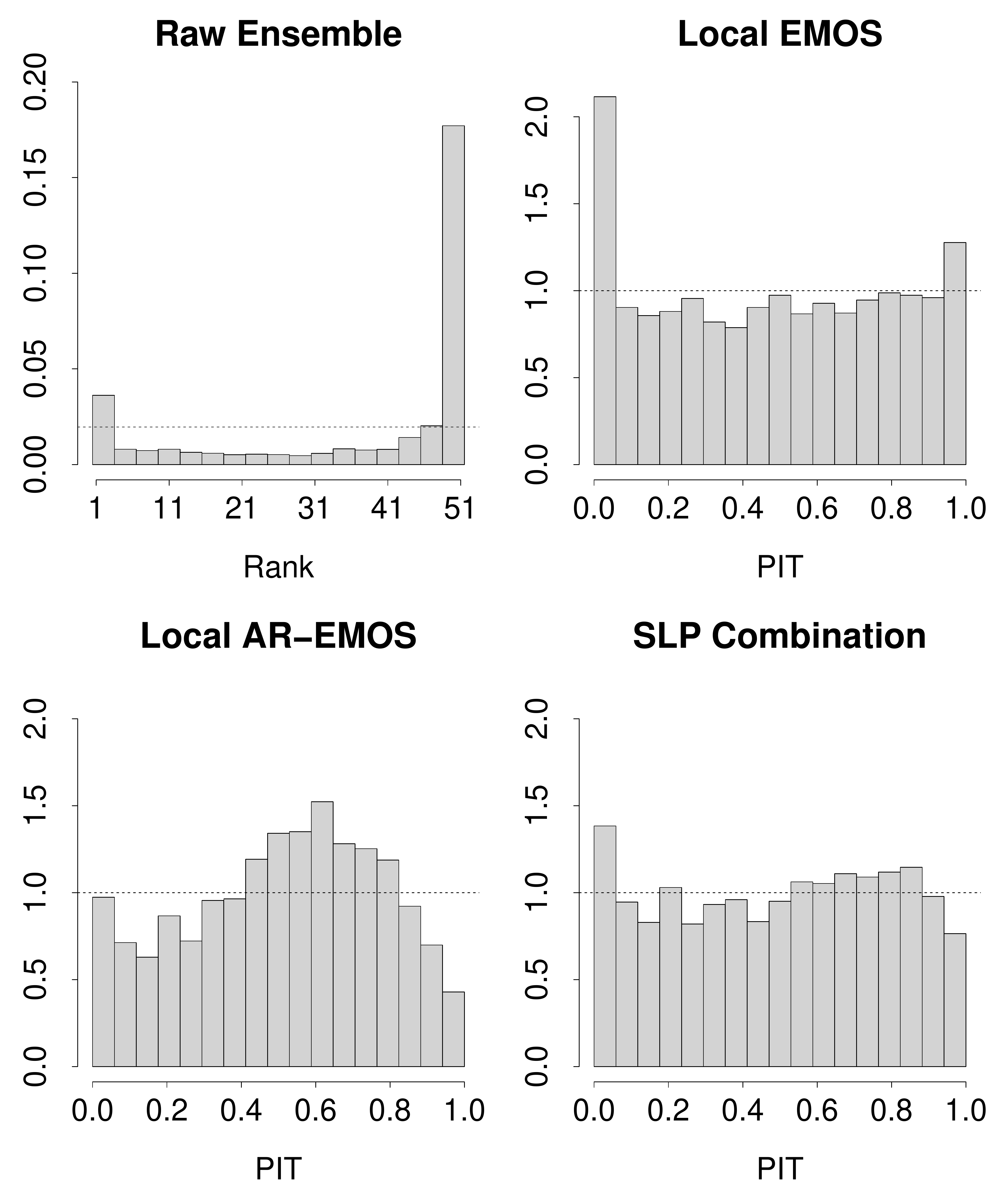}\\
  \caption{Univariate verification rank histogram and PIT histograms of local EMOS, local AR-EMOS, and SLP combination
  (with $w_{1}= w_{2} =0.5$, $c=0.9$) over 3650 days for station Frankfurt a.M.}\label{f2}
\end{figure}

\section{Application to ECMWF data at a single station} \label{sec:applicationsingle}

The above verifying results are obtained by averaging over all stations used for the case study. Since the
discussed methods refer to a stations-wise approach, it should
also be of interest to investigate their performance with respect to a single station. For this analysis we use the 50 member ECMWF forecast ensemble and SYNOP observations as well. However, we employ a data base containing about 10 years and ranging from dates in 2002 up to 2012, which is part of a larger data set investigated in \citet{Hemri&2014}.
To this end, the station Frankfurt a.M. in the south of Germany is exemplarily observed during a verification period
from 2002-04-27 to 2012-04-23 (3650 days).

\begin{table}
\caption{Order $p$ of autoregressive fit for 3650 forecast days at Frankfurt a.M.}\label{t3}
\centering
\begin{tabular}{lcccccc}
\hline
$p$ & 0 & 1 & 2 & 3 & 4 & 5\ldots 15\\
\hline
Freq. & 1314 & 1443 & 391 & 130 & 197 &  175\\
\hline
\end{tabular}
\end{table}

\subsection{Investigation of autoregressive fit}

For illustrative purposes, the AR modification introduced in Section \ref{sec:ar} is applied to the raw ensemble mean
of each forecast day based on a training period of $T_{1} = 90$ previous days.
Table \ref{t3} displays the absolute frequencies of the
order $p$ chosen by the minimal AIC criterion for an autoregressive fit to the error series.
Although 1314 out of 3650 of these series used for prediction do
reveal no substantial autocorrelation, thus yielding a simple AR(0) fit, the majority of cases exhibits an autoregressive structure that needs to be accounted for. The by far most prominent type of autoregressive structure is the AR(1) process, fitted to the error series in 1443 cases. Even higher orders of $p=2$, $p=3$ or $p=4$ appear quite frequently. Only for orders of $p=5$ or higher, the number of cases is decreasing.
This observation is in line with other applications of AR processes (in econometric as well as in environmental/meteorological applications like those briefly discussed in Section \ref{sec:intro}), most of the dependence structures present in data can be covered by autoregressive processes of low orders (e.g. $p=1$ or $p=2$).

The analysis of the orders $p$ fit to the error series reveals the usefulness and flexibility of our proposed method. It is able to adapt to the type of autoregressive dependence of the errors. In cases where there is a substantial autocorrelation present in the series, the AR-modification method chooses an appropriate order and corrects the ensemble for the autoregressive structure by performing an AR-modified bias correction. In cases where there is no substantial autocorrelation, our method performs a simple bias correction based on the past $T_1$ values of the forecast errors.

\subsection{Combination of predictive distributions}

By proceeding along the same lines as in Section \ref{sec:applicationtemp},
we set up the local AR-EMOS predictive distribution based on applying the AR-modification to the raw ensemble as in Section \ref{sec:preddistoverall}, and in a second step obtain the SLP combination of local EMOS and local AR-EMOS as described in Section \ref{sec:slpoverall}.

We investigated the same grid of 99 values for $w_1$ and $c$ that was considered in Section \ref{sec:slpoverall} to obtain the optimal SLP parameters with respect to the CRPS.
As it turns out, the combination with
weights $w_{1} = w_{2} = 0.5$ and spread-adjustment $c=0.9$ performs best with respect to
a minimal CRPS.

\begin{table}
\caption{PIT variance (dispersion) and density forecast root mean variance (sharpness)
for station Frankfurt a.M. for $3650$ verification days.}\label{t5}
\centering
\begin{tabular}{lll}
\hline
& $\text{Var(PIT)}$ & $\text{RMV}$ \\
\hline
Local EMOS & 0.098 & 1.50\\
Local AR-EMOS & 0.068 & 1.82\\
SLP Combination ($w_{1} =0.5, c=0.9$) & 0.086 & 1.54\\
\hline
\end{tabular}
\end{table}

Figure \ref{f2} shows the verification rank histogram of the raw ensemble and the PIT histograms of the three methods over 3650 days, behaving similar
to the PIT histograms computed over all 383 stations in Figure \ref{f1}.
Particularly the raw ensemble and local EMOS exhibit an additional forecast bias. While the raw ensemble tends to underestimate the temperatures at Frankfurt a.M., local EMOS tends to overestimate it. The hump shape of the local AR-EMOS predictive distribution already seen in Figure \ref{f1}, is even more pronounced when investigating the PIT values of a single station. As in Figure \ref{f1}, our SLP combination is closest to uniformity, although the first bin is more occupied than in the overall PIT histogram of Figure \ref{f1}.
These observations are in line with the results of  Table \ref{t5}, showing the variance of the PIT values and the root mean variance as a measure of sharpness. Similar to the situation presented in the left panel, the variance of the local EMOS PIT values is larger than $\frac{1}{12}$, indicating underdispersion, while the variance of the local AR-EMOS PIT values is smaller than $\frac{1}{12}$, indicating overdispersion. When considering the PIT values only for the station Frankfurt a.M., the variance of the PIT values is even smaller than in the overall case (Table \ref{t4}), corresponding to the more pronounced hump-shape.
The variance of the PIT values of the SLP combination differs only slightly from $\frac{1}{12}$, indicating that the PIT histogram is close to uniformity.
The sharpness properties of the predictive distribution directly correspond to those in Table \ref{t4}. While local EMOS is sharpest and local AR-EMOS is least sharpest, our SLP combination has a sharpness value between the other two.

The Diebold-Mariano test statistic introduced in Section \ref{sec:dmtest}
for comparing the predictive performance of local EMOS with our derived SLP combination (here applied to the CRPS series at the station Frankfurt)
is $S=9.4624$ for $h=1$, thereby indicating a highly significant
nonequal predictive performance of the two methods.

\section{Concluding Remarks} \label{sec:discussion}

In this work, we propose a basic AR-modification method that accounts for potential autoregressive structures in forecast errors of the raw ensemble.
The AR-modification is straightforward to compute by standard \verb|R| functions to fit AR processes and can be utilized in the context of ensemble forecasts in different ways. It can simply be employed to obtain an adjusted ''raw'' ensemble of forecasts that is corrected for autoregressive structures or it can be used to construct different types of predictive distributions.

To this end, our proposed modification is simple and yet effective.
In our case study we suggest to built an EMOS-like predictive distribution based on the AR-adjusted ensemble and in a second step obtain an aggregated predictive distribution that comprises of the state-of-the-art EMOS predictive distribution and our AR-EMOS variant.
As EMOS is a well-established standard postprocessing procedure that is easy to compute, our proposed extension can be easily constructed. Further, the approach is neither restricted to a specific postprocessing model nor to temperature forecasts.
A modification of the approach allowing to combine the method with other standard postprocessing models such as BMA should be straightforward.
Our AR-modification approach allows for a flexible bias-correction based on fitting AR process to the error series. The method identifies cases where a correction for autoregressive structures is indicated and performs a simple bias-correction based on past values in cases where no substantial autocorrelation is present.
In the considered case study all our derived variants based on the AR-modification approach improve on the standard EMOS method. While the improvement of the AR-EMOS predictive distribution on the standard EMOS distribution is only small and AR-EMOS still lacks calibration, the SLP combined predictive distribution based on EMOS and AR-EMOS shows a PIT histogram close to uniformity and the improvement on standard EMOS with respect to the verification scores is highly significant.

In line with the recently increased interest in multivariate postprocessing models that yield physically coherent forecasts, it should be of interest to extend our method to this field of research. An approach that allows to retain dependence structures with low computational cost is the Ensemble Copula Coupling (ECC) method introduced by \citet{Schefzik&2013}. ECC is able to recover temporal, spatial and inter-variable dependencies present in the raw ensemble by reordering samples from the predictive distributions according to the rank structure of the raw ensemble. It is a flexible and computationally efficient method, as one needs to compute simply the rank order structure of the raw ensemble. A combination e.g. of our SLP predictive distribution with ECC would be straightforward and easy to compute. This procedure can account for spatial dependence structures between the stations not considered by our station-wise approach and it may even be able to recover additional temporal dependencies from the raw ensemble, that are not explicitly modelled by our AR-approach that only considers the autoregressive structure in the errors.

While ECC is a nonparametric method that can recover different types of multivariate structures simultaneously, there are several parametric approaches to incorporate spatial or inter-variable dependencies.
\citet{Berrocal&2007} and \citet{Kleiber&a2011} for example propose spatially adaptive extensions of the basic BMA method, while \citet{ScheuererBuermann2014}, \citet{ScheuererKoenig2014} and \citet{Feldmann&2015} investigate different ways of extending EMOS to incorporate spatial dependencies.
There is also an interest to investigate inter-variable dependencies. For example \citet{Schuhen&2012} develop a bivariate EMOS model for wind vectors and \citet{baran2015joint} a bivariate BMA model for temperature and wind speed.
Further, \citet{moller2013multivariate} investigate a general multivariate setting that allows to combine arbitrary univariate postprocessing distributions within a Gaussian copula framework.

The development of multivariate postprocessing models is a very active area of research, and an extension of our proposed AR-modification that incorporates spatial or inter-variable dependencies as well should be highly beneficial.

\section*{Acknowledgements}
We are grateful to the European Centre for Medium-Range
Weather Forecasts (ECMWF) and the German Weather Service (DWD)
for providing forecast and observation data, respectively. We wish to thank Tilmann Gneiting for useful discussions and helpful comments.

\end{document}